\documentclass{iau} 
\usepackage{graphicx}
\usepackage{natbib}
\usepackage{amsmath}
\usepackage{caption}
\usepackage{subcaption}

\pubyear{2022}
\volume{370}
\jname{Winds of Stars and Exoplanets}
\editors{A.~A. Vidotto, L.~Fossati \& J.~Vink, eds.}

\title[DACs and local magnetic spots] %% give here short title %%
{Discrete Absorption Components from 3-D spot models of hot star winds}

\author[F.~A.~Driessen and N.~D.~Kee]   %% give here short author list %%
{F.~A.~Driessen$^{1,2}$, N.~D.~Kee$^2$}

\affiliation{$^1$Institute of Astronomy, KU Leuven, Celestijnenlaan 200D/2401, 3001 Leuven, Belgium \\ [\affilskip]
 $^2$National Solar Observatory, 22 Ohi`a Ku St., Makawao, HI 96768, USA}

\renewcommand{\vec}[1]{\ensuremath{\boldsymbol{#1}}}

\begin{document}

\maketitle

\begin{abstract}
The winds of hot, massive stars are variable from processes happening on both large and small spatial scales. A particular case of such wind variability is ‘discrete-absorption components’ (DACs) that manifest themselves as outward moving density features in UV resonance line spectra. Such DACs are believed to be caused by large-scale spiral-shaped density structures in the stellar wind. We consider novel 3-D radiation-hydrodynamic models of rotating hot star winds and study the emergence of co-rotating spiral structures due to a local (pseudo-)magnetic spot on the stellar surface. Subsequently, the hydrodynamic models are used to retrieve DAC spectral signatures in synthetic UV spectra created from a 3-D short-characteristics radiative transfer code.

%% add here a maximum of 10 keywords, to be taken from the file <Keywords.txt>
\end{abstract}

\firstsection % if your document starts with a section,
              
\section{Introduction}
%The text of your contribution starts here. You can use using the cite command like this: \cite{2014MNRAS.441.2361V}. A sample figure is shown below.

Line-driven wind outflows from hot, massive OB-stars are mainly caused by stellar photons scattering off metallic ions in the stellar atmosphere \citep{1975ApJ...195..157C}. However, instead of being homogeneous the wind outflow is rather structured on small- and large-spatial scales. A type of large-scale \emph{coherent} wind structure is believed to originate from spiral arms akin to co-rotating interactions regions (CIRs) in the solar wind \citep{1986A&A...165..157M} whereby the spiral is initiated close to the stellar surface \citep[e.g.][]{2015ApJ...809...12M}. The origin of such spiral structures in hot star winds remains, however, illusive with non-radial stellar pulsations \citep{2008ApJ...678..408L}, local stellar spots \citep{1996ApJ...462..469C,2017MNRAS.470.3672D}, or potentially both mechanisms together as potential causes. More recently, the idea of short-lived stellar prominences has been put forward as an explanation of spiral structures \citep{2016A&A...594A..56S}.

Particularly, CIRs in line-driven winds have gained considerable attention as they may explain observed line-profile variability in UV resonance lines among the sample of OB-stars \citep{1989ApJS...69..527H}. This variability presents as `discrete absorption components' (DACs), which represent localised absorption features that move from line centre toward line edge. The time needed to complete such migration typically amounts to a day up to a few days, a fact which appears to be intimately correlated with the (projected) stellar rotation. Therefore, there is strong believe that DACs are rotationally modulated \citep[e.g.][and references therein]{1999A&A...344..231K}. 

To model line-driven wind CIRs and their DAC signature we here present 3-D radiation-hydrodynamic models of a line-driven wind from a typical O-star in our Galaxy. We take into account the effects of a small-scale sub-surface magnetic field and its influence on the wind outflow. Such small-scale, localised magnetic fields are thought to arise from the spatially restricted sub-surface iron-recombination convection zone of OB-stars \citep{2011A&A...534A.140C} leading to the formation of bright star spots.

\section{Magnetic spot model}

We model the local emergence of a hot star spot from the iron-recombination zone by considering at the stellar surface a pure vertical magnetic field of strength $B$ in magnetohydrostatic equilibrium with its surroundings. Horizontal pressure balance dictates that the sum of gas and magnetic pressure inside the spot must balance the pressure of the ambient medium: $P_\mathrm{spot} + P_\mathrm{mag} = P_\mathrm{a}$. Invoking the ideal gas law, $P=\rho k_\mathrm{B}T/\mu$, it follows that the density inside the spot $\rho_\mathrm{s}$ is lower compared to the density outside the spot
\begin{equation}
\rho_\mathrm{spot} = \rho_\mathrm{a} - \frac{P_\mathrm{mag}}{k_\mathrm{B}T/\mu} = \rho_\mathrm{a} - \frac{B_\mathrm{s}^2}{8\pi k_\mathrm{B}T/\mu}.
\end{equation}

To include the effect of the spot on the radiation force in our model we consider the spot to be located in a gray radiative atmosphere. At a given optical depth $\tau$ we may then write the atmospheric temperature stratification as
\begin{equation}
T^4(\tau) = 0.75 T_\mathrm{eff}^4 ( \tau + 2/3 ),
\end{equation}
and the local optical depth above some height $z$ can be expressed as
\begin{equation}
\tau(z) \equiv \int_z^{+\infty} \kappa(z') \rho(z')\,dz' = \kappa \rho H,
\end{equation}
where the last equality assumes a spatially constant opacity $\kappa$ inside the stratified atmosphere with barometric scale height $H=k_\mathrm{B}T/(\mu g)$, and $g$ the local gravity. By combining the above relations, it follows that the spot optical depth
\begin{equation}
\tau_\mathrm{spot} = \tau_\mathrm{a} - \frac{\kappa H B_\mathrm{spot}^2}{8\pi k_\mathrm{B}T/\mu} = \tau_\mathrm{a} - \frac{\kappa B_\mathrm{spot}^2}{8\pi g}.
\end{equation}
At the height where the spot reaches optical depth $\tau_\mathrm{spot}=2/3$---often defined as the visible surface---the ambient atmosphere has a higher optical depth, $\tau_\mathrm{a} = 2/3 + \kappa B_\mathrm{spot}^2/(8\pi g)$, such that the star spot appears \emph{bright} relative to its surroundings.

The brightness of the spot will perturb the radiation flux to be higher inside the spot $F_\mathrm{spot}$ relative to the unperturbed stellar flux $F_\star$. If we assume that the spot size is small to not distort the overall atmospheric temperature structure the radiative flux enhancement from the spot can be expressed as
\begin{equation}\label{eq:fspot}
\frac{F_\mathrm{spot}}{F_\star} = \left( \frac{T(\tau_\mathrm{spot}=2/3)}{T(\tau_\mathrm{a}=2/3)} \right)^4 = 1 + \frac{3}{4}\frac{\kappa B_\mathrm{spot}^2}{8\pi g} = 1 + 0.5\left(\frac{B_\mathrm{spot}}{B_\mathrm{crit}} \right)^2,
\end{equation}
where $B_\mathrm{crit}^2 \equiv 16\pi g/(3\kappa)$ is a critical magnetic field strength for which the magnetic pressure equals to the photospheric gas pressure at the $\tau =2/3$ surface. For the case that $B_\mathrm{crit} = B_\mathrm{spot}$ the small spot will provide a 50\% increased amplification in the radiation force compared to the unperturbed stellar radiation force.

\section{Wind dynamics from radiation-hydrodynamic simulations}

The above (local) spot flux enhancement can be readily incorporated in a radiation-hydrodynamic simulation of a line-driven wind outflow. To that end we solve the 3-D spherically-symmetric hydrodynamic equations for a proto-typical Galactic O-supergiant star with parameters $L_\star=9\times 10^5 L_\odot$, $M_\star = 50M_\odot$, $R_\star=20R_\odot$, and $T_\mathrm{eff}=40\,000$\,K. The star is moderately rotating ($v_\mathrm{rot}\approx 100$\,km/s) at 20\% of its critical rotation speed. Such slow rotation relieves us from considering more involved effects like gravity darkening and taking into account the oblateness of the star \citep{1996ApJ...462..469C}. Finally, we assume an isothermal wind outflow, which is justifiable with the high densities inside the wind that make radiative cooling very efficient 

We adopt the Sobolev approximation for the line-driven wind \citep{1975ApJ...195..157C} meaning that we suppress the effects of a powerful radiation-triggered instability in the wind \citep[e.g.][and references therein]{2022A&A...663A..40D}. Within this Ansatz the \emph{vector} radiation line force is computed by performing at every wind point a quadrature over the solid angle, $d\varOmega = \sin \varTheta \,d\varTheta \,d\varPhi$, of the stellar core intensity for a set of rays in direction $\vec{\hat{n}}$, weighted by the line-of-sight velocity gradient:
\begin{equation}
\vec{g}_\mathrm{line}(\vec{r}) = \frac{(\kappa_\mathrm{e} \bar{Q})^{1-\alpha}}{(1-\alpha)c^{\alpha+1}} \oint_\varOmega \,d\varOmega\, \vec{\hat{n}}\,I_\star(\vec{\hat{n}}) \left[ \vec{\hat{n}}\cdot \nabla(\vec{\hat{n}}\cdot \vec{v}) /\rho(\vec{r}) \right]^\alpha,
\end{equation}
which is added as a source term to the hydrodynamic momentum equation. Here $\kappa_\mathrm{e}=0.34$\,cm$^2$/g is the free-electron scattering opacity and $I_\star$ is the unattenuated stellar core intensity that is ray-by-ray weighted according to where the radiation emerges on the surface---for instance to account for limb darkening. The quantity $\alpha$ is the CAK power-law index and $\bar{Q}$ sets the ensemble-integrated line strength \citep[see][for more details on these atomic quantities]{2022arXiv220409981P}. These CAK line-ensemble distribution parameters are taken to be the standard O-star Galactic values of $\alpha=0.65$ and $\bar{Q}=2000$. Note that along a given ray the line-of-sight velocity gradient tensor $\vec{\hat{n}}\cdot \nabla(\vec{\hat{n}}\cdot \vec{v})$ depends exclusively on local quantities and can be readily computed. Initial conditions are set from a smooth CAK wind while the stellar boundary adopts a constant mass density with the radial velocity extrapolated into the boundary, the polar velocity is set under a no-slip condition, and the azimuthal velocity is set to the stellar rotation. Polar boundaries are set by symmetric/asymmetric combinations and in azimuth we apply periodic boundary conditions.

The required angle quadrature is evaluated by distributing radiation rays across the stellar disc. Experimentation has shown that a ray quadrature of $(n_\varTheta, n_\varPhi)=(6,6)$ rays is sufficient to model the overall wind dynamics. For the radiation polar angle $\varTheta$ the ray emerging positions and flux weights are computed with a Gauss--Legendre quadrature. For the radiation azimuthal angle $\varPhi$ the ray emerging positions and flux weights are taken uniformly distributed across the full $2\pi$ projected stellar disc. Furthermore points with the same $\Theta$ have their $\Phi$ adjusted by $\pm 20^\circ$ to better sample the full stellar disc. In case a ray is emerging from inside the bright spot, we increase the ray flux weight by an amount set through Eq.~\eqref{eq:fspot} (for our chosen star $B_\mathrm{crit}\approx 300$\,G). The spot has a size of $10^\circ$ and has a Gaussian decay of field strength towards its edge. In analogy with previous line-driven CIR models, we simulate a half hemisphere, such that over the entire star two symmetric spots appear.

\begin{figure}[t]
     \centering
     \begin{subfigure}[b]{0.65\textwidth}
         \centering
         \includegraphics[width=\textwidth]{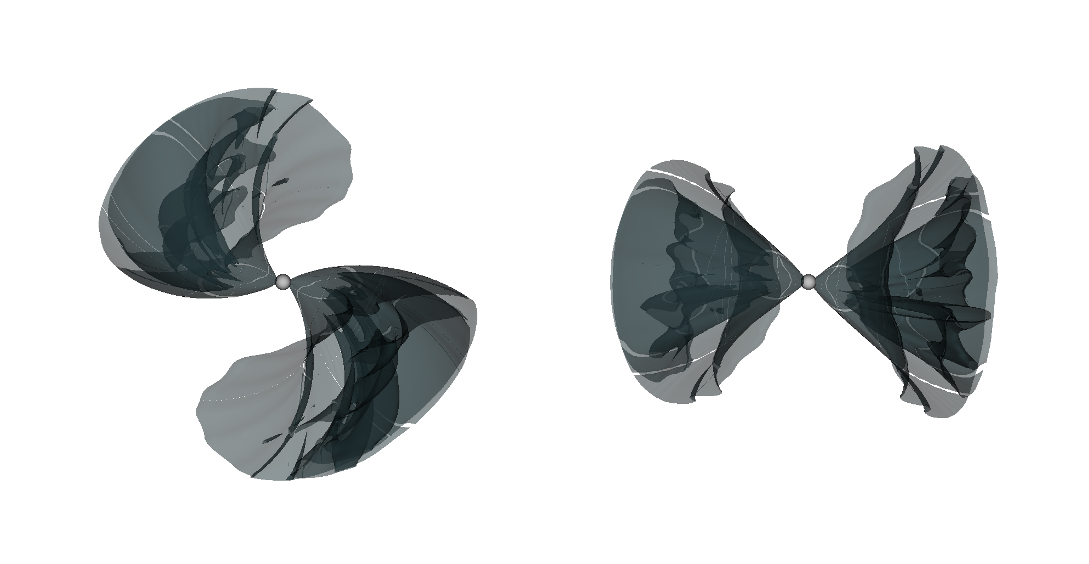}
         \caption{3-D isocontours showing wind density values 1.5\% above the mean wind mass density. The displayed wind dynamics results from a pseudo-magnetic spot of strength 1000\,G. \emph{Left}: Line-of-sight structure seen under $45^\circ$ from the pole. \emph{Right}: Line-of-sight structure under $90^\circ$ from the pole. Centred is the solid, grey star.}
         \label{fig:3d_1000g}
     \end{subfigure}
     \vfill
     \begin{subfigure}[b]{0.65\textwidth}
         \centering
         \includegraphics[width=\textwidth]{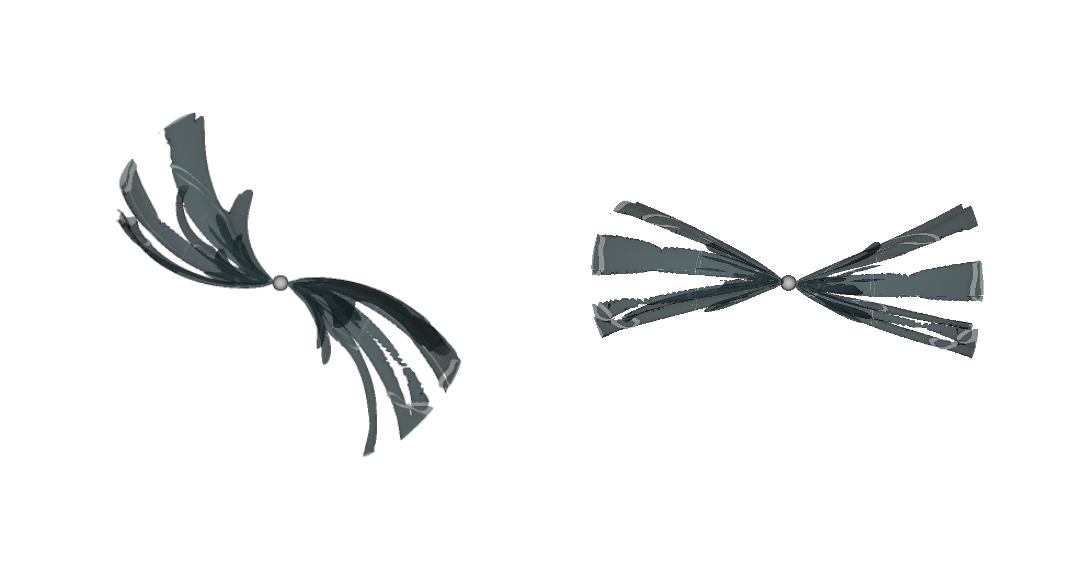}
         \caption{Same as in (a), but for a spot of strength 150\,G. We note that the weak magnetic field model generates an overall weak CIR such that it is challenging to visualise its structure.}
         \label{fig:3d_150g}
     \end{subfigure}
\end{figure}

In Figures \ref{fig:3d_1000g} and \ref{fig:3d_150g} we display a set of 3-D isodensity contours from the resulting (pseudo-)magnetic spot and wind interaction for both of the magnetic field strengths (150\,G and 1000\,G). Although such 3-D models lend themselves well to display the overall wind dynamics, they are somewhat harder for the quantitative interpretation due to the visualisation. Nonetheless, when we consider 2-D cuts in the meridional and equatorial plane we find overall wind dynamics and properties akin to previous 2-D line-driven wind models aiming to model line-driven CIRs \citep[see][for extensive discussions]{1996ApJ...462..469C,2017MNRAS.470.3672D}. Notable differences in the 3-D wind dynamics concern the fact that the additional flow direction weakens compressions such that CIRs appear less strong compared to 2-D models. Moreover, the azimuthal line force component can yield a net spin-up of wind material ahead of the spot with a spin-down in regions trailing the spot \citep{2000ApJ...537..461G}. Overall this effect azimuthally broadens the 3-D CIR compared to its 2-D analogue while also softening the effect of the spot. Finally, an effect intrinsic to the 3-D models is the fact that all wind material pushed by the spot flux is contained within the cone angle covered by the spot.

It should be noted that these previous 2-D line-driven CIR models only performed ad-hoc flux enhancements of some putative star spot---although \citet{2017MNRAS.470.3672D} aim to more realistically model possible spot flux enhancements using spectroscopic constraints. This is in contrast with the present models wherein a first effect, albeit crude and simplified, of a (pseudo-)magnetic spot is taken into account. The models displayed in Figure \ref{fig:3d_1000g} and \ref{fig:3d_150g} also serve as a direct input to the 3-D short-characteristics radiative transfer computation discussed in the next section. This allows future comparison of constraints on the flux enhancement and DAC strength to limits on the unobserved, localised patches of stellar magnetic field.

\section{Synthetic time series of a proto-typical UV resonance line}

\begin{figure}[t]
     \centering
     \begin{subfigure}[b]{0.65\textwidth}
         \centering
         \includegraphics[width=\textwidth]{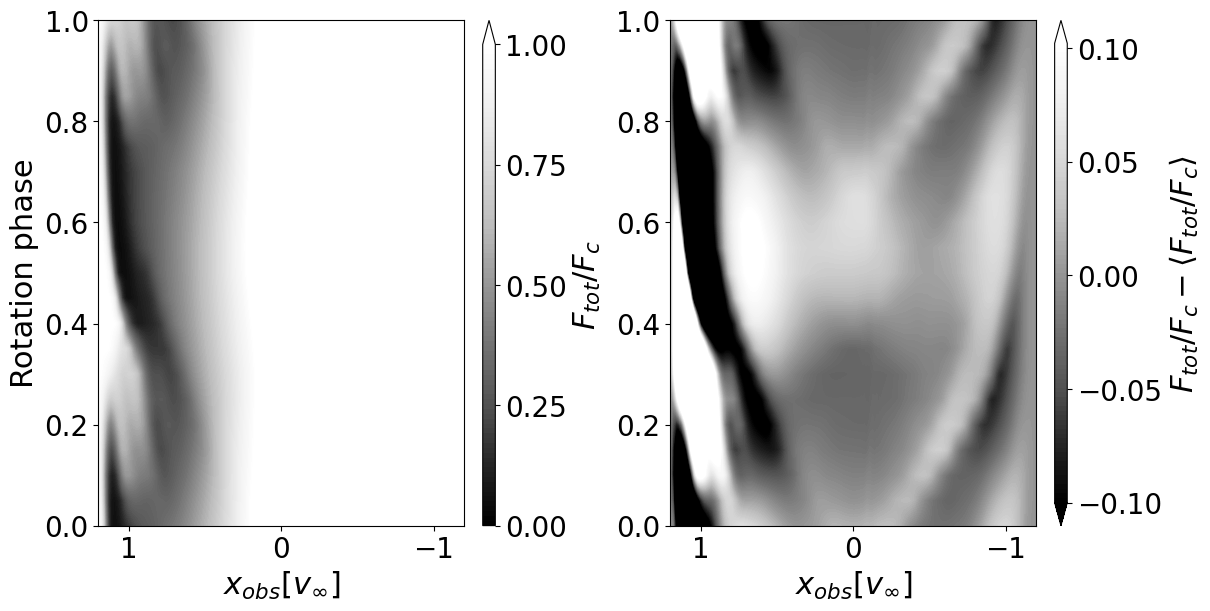}
         \caption{Temporal variation of the \textup{C\,\textsc{\lowercase{iv}}} UV resonance line for a pseudo-magnetic spot of strength 1000\,G. \emph{Left}: Flux with respect to the continuum. \emph{Right}: Temporally averaged normalised spectrum subtracted from the line profile in the left panel with variations limited to $\pm 10$\% with respect to the mean spectrum. This serves to better show the line-profile variability.}
         \label{fig:dac_1000g}
     \end{subfigure}
     \vfill
     \begin{subfigure}[b]{0.65\textwidth}
         \centering
         \includegraphics[width=\textwidth]{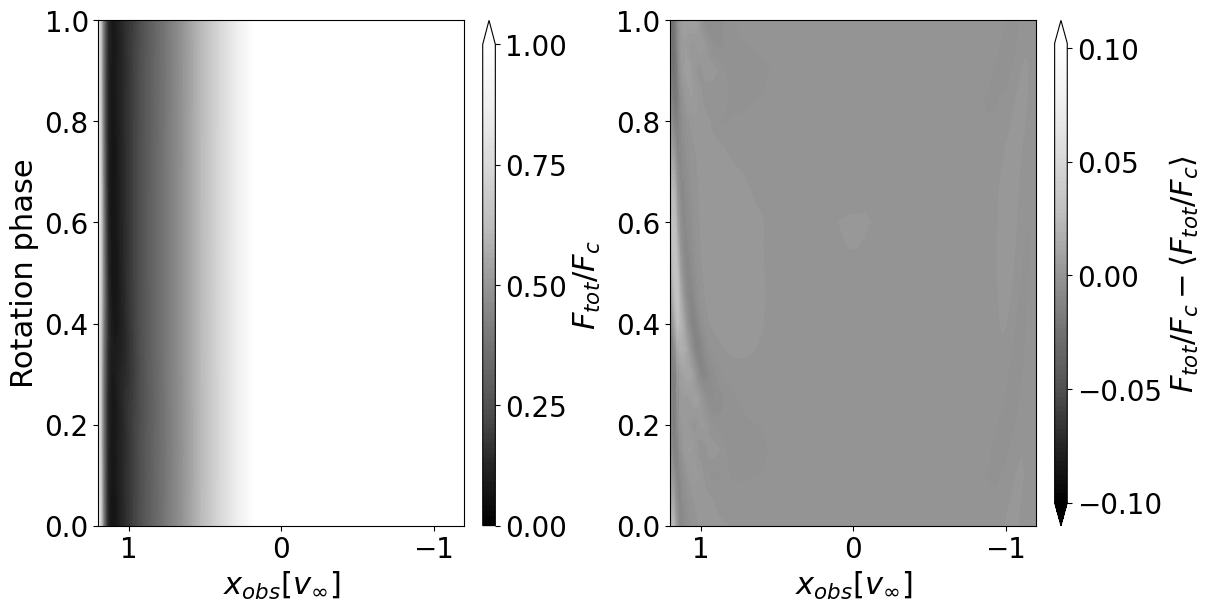}
         \caption{Same as in (a), but for a spot of strength 150\,G.}
         \label{fig:dac_150g}
     \end{subfigure}
\end{figure}

Using the wind structure from our radiation-hydrodynamic simulations, Figures \ref{fig:dac_1000g} and \ref{fig:dac_150g} display the resulting synthetic UV resonance line spectrum of \textup{C\,\textsc{\lowercase{iv}}} obtained from solving the 3-D radiation transfer problem with the method of short-characteristics \citep[see][for full details]{2020A&A...633A..16H}. 

The time series of the 1000\,G spot model indeed \emph{quantitatively} reproduce the expected DAC shape---we stress that DAC signatures are star-dependent in their shape and duration. Indeed, excess absorption occurs in the blue-edge of the line profile starting from velocities $\approx 0.5v_\infty$. Since line-driven winds already attain supersonic speeds about 1\% above the stellar surface and $v\approx 0.5v_\infty$ around $1R_\star$ above the surface, this shows that the CIR creating the DAC should form close to the star \citep{2015ApJ...809...12M}. The rather `broad' absorption signal then gradually moves outward becoming narrow near the blue edge of the line at terminal wind speeds. This can be understood from the fact that near the star a collection of gas parcels, that have the correct optical depth to absorb, also possess a large range of velocity dispersion in the dense inner star region. This leads to a broad initial absorption near $v\approx 0.5v_\infty$. On the contrary, gas parcels that absorb at the blue edge populate the outer edges of the winding spiral arm and are generally all moving at or near the terminal wind speed. This means these gas parcels posses much less velocity dispersion, hence the DAC absorption excess narrows farther away from the star at high wind speeds.

This general behaviour is in contrast to the model with a spot field strength of 150\,G. Here a DAC absorption signature is only modestly visible in the right-hand panel of Figure \ref{fig:dac_150g} that subtracts the mean spectrum from all spectra. Intuitively one may expect this is due to the lack of density enhancement in the spiral arm (see Figure \ref{fig:3d_150g}), but in fact the generation of DACs in line-driven CIRs is mainly dependent on kinks in the \emph{velocity field} \citep{1996ApJ...462..469C}. The kinks in the velocity field are, however, also weak due to the rather modest flux enhancement the spot generates. As such no significant DAC feature can form. Caution is warranted since the flux enhancement predicted by Eq.~\eqref{eq:fspot} depends on free parameters, notably, the assumed mass-absorption coefficient $\kappa$ in the atmosphere. Since a 150\,G magnetic field strength is expected to arise from a dynamo-generated spot \citep{2011A&A...534A.140C}, we have performed an additional spot model (not shown here) employing a less conservative value for the opacity, $\kappa=\kappa_\mathrm{OPAL}\approx 1$, near the iron-recombination zone \citep[][their Fig.~1]{2009A&A...499..279C}. Adopting this opacity lowers the critical field strength to $B_\mathrm{crit}\approx 170$\,G and yields larger spot flux enhancement. Nonetheless, our simulations predict only a marginally denser, compressed CIR with a modest amount of extra absorption in the blue edge of the DAC compared to Figure \ref{fig:3d_150g}. A potential reason for this behaviour at low spot field strengths is that for the assumed spots in this work of $10^\circ$ in size, the Gaussian decay applied to the spot field strength `shuts off' too fast the flux enhancement over the geometrical extent the spot can contribute to. Future investigation of characteristic spot properties is thus called for.

\section{Conclusion}

We find that a bright star spot with a strong magnetic field strength, typically associated with the inferred large-scale field strengths of OB-stars, produces a strong enough absorption signature and expected DAC shape. On the other hand, a spot with a magnetic field strength predicted to arise from a dynamo operating in the sub-surface convection zone \citep{2020ApJ...900..113J} produces almost no absorption features in the DAC. \emph{A word of caution is in place}, however, in the interpretation of these results given the model uncertainties involved. To assess whether magnetic spots can generate line-driven CIRs and explain DACs, full radiation-\emph{magneto}hydrodynamic models are necessary. These models would need improved input physics on the spot size, spot magnetic field strength (strong vs.~weak), the local field geometry (simple loops vs.~complex topology), the lifetime of the spot (short vs.~long compared to stellar rotation), and/or the rotation properties of the spot (co-rotating vs.~differentially rotating). This may then solve some of the puzzling outcomes found for the weak field strength spot model here. Future theoretical and observational efforts are thus further required to assess the link between small-scale surface magnetic fields in OB-stars and DAC signatures.

%\acknowledgment{F.A.D acknowledges funding for a research stay abroad at the National Solar Observatory (NSO) in Hawai‘i provided by the Research Foundation - Flanders (FWO) under grant K204122N. F.A.D is grateful for the generous hospitality offered at NSO to conduct this work, particularly by its director, Dr.~Thomas Rimmele.}

%Definition of a few common journal names 
\def\apj{{ApJ}}    
\def\nat{{Nature}}    
\def\jgr{{JGR}}    
\def\apjl{{ApJ Letters}}    
\def\aap{{A\&A}}   
\def\mnras{{MNRAS}}
\def\aj{{AJ}}
\let\mnrasl=\mnras

%Alternatively, one can use natbib linking to their own bibtex library
%%
%\bibliographystyle{aa}
%\bibliography{my-own-bibtex-list}

\end{document}